\documentclass[11pt,a4paper]{article}

\usepackage[T1]{fontenc}
\usepackage[utf8]{inputenc}
\usepackage[english]{babel}
\usepackage{lmodern}
\usepackage{amsmath,amssymb,amsthm}
\usepackage[margin=25mm]{geometry}
\usepackage[hidelinks]{hyperref}
\hypersetup{
  pdftitle={Minimal Input Dimension for Superactivation of One-Shot Zero-Error Quantum Capacity},
  pdfauthor={A. V. Shcherbakov},
  pdfsubject={One-shot zero-error quantum capacity},
  pdfkeywords={one-shot zero-error quantum capacity, superactivation, noncommutative graphs, quantum channels, additivity}
}
\newcommand{\C}{\mathbb C}
\newcommand{\ket}[1]{\lvert #1\rangle}
\newcommand{\bra}[1]{\langle #1\rvert}
\newcommand{\spn}{\operatorname{span}}
\newcommand{\diag}{\operatorname{diag}}
\newcommand{\rank}{\operatorname{rank}}
\newcommand{\Ran}{\operatorname{Ran}}
\newtheorem{lemma}{Lemma}
\newtheorem{theorem}{Theorem}
\newtheorem{proposition}{Proposition}
\newtheorem{corollary}{Corollary}

\title{Minimal Input Dimension for Superactivation of\\
One-Shot Zero-Error Quantum Capacity}
\author{A.~V.~Shcherbakov\thanks{\href{mailto:shcherbakov.av@phystech.edu}{\nolinkurl{shcherbakov.av@phystech.edu}}}\\[0.5em]
\small Steklov Mathematical Institute of the Russian Academy of Sciences\\
\small Moscow, Russia}
\date{}

\begin{document}
\maketitle
\begin{abstract}
\normalsize
The one-shot zero-error quantum capacity $\overline{Q}_0$ is studied within the framework of noncommutative graphs of quantum channels. The additivity of $\overline{Q}_0$ is proved for an arbitrary finite collection of channels with two-dimensional input spaces. An explicit pair of noncommutative graphs $S,T\subset M_3$ exhibiting superactivation of the one-shot zero-error quantum capacity is also constructed. Together, these results establish that the minimum input dimension for such superactivation is $3$.
\end{abstract}
\noindent\textbf{Keywords:} one-shot zero-error quantum capacity, superactivation, noncommutative graphs, quantum channels, additivity.

\section{Introduction}
Several different notions of capacity arise in the theory of quantum channels, since the same physical channel can transmit different information resources and the communication protocol can be subject to different constraints. A detailed account of the basic definitions can be found, for example, in \cite{holevo}.

The quantum capacity $Q(\Phi)$ describes the asymptotic rate of reliable transmission of quantum states. Coherent information plays a central role in its theory \cite{lloyd,devetak}. Quantum capacity is not additive under tensor products of channels: there exist two channels whose individual quantum capacities vanish, while their tensor product has positive quantum capacity \cite{smith}. This effect is known as superactivation.

A separate class of problems arises when the error is required to vanish exactly, rather than asymptotically. This paper considers the transmission of quantum states using a channel once; the corresponding one-shot zero-error quantum capacity is denoted by $\overline{Q}_0$. Its asymptotic counterpart is denoted by $Q_0$:

\[
Q_0(\Phi)=\sup_{n\ge1}\frac{1}{n}\overline{Q}_0(\Phi^{\otimes n}).
\]

Noncommutative graphs, introduced for example in \cite{duan}, provide a natural language for describing such problems.

We now give precise definitions. All Hilbert spaces considered here are finite-dimensional and complex. Let $B(H)$ denote the algebra of linear operators on $H$, and let $\mathcal S(H)=\{\rho\in B(H):\rho\ge0,\ \operatorname{Tr}\rho=1\}$ denote the set of states. Set $M_r=B(\C^r)$, the algebra of complex $r\times r$ matrices, and let $I_r$ be its identity matrix.

For a quantum channel
\[
\Phi:\mathcal S(H_A)\longrightarrow\mathcal S(H_B),
\]
with a Kraus representation
\[
\Phi(\rho)=\sum_k V_k\rho V_k^\dagger,
\qquad \sum_k V_k^\dagger V_k=I_{H_A},
\]
the noncommutative graph of the channel is the operator system \cite{duan}
\[
S(\Phi)=\spn\{V_k^\dagger V_l:k,l\}\subseteq B(H_A).
\]
It contains the identity and is closed under Hermitian conjugation. In general, an operator system is a complex linear subspace $S\subseteq B(H)$ such that $I_H\in S$ and $S=S^\dagger$.

For a tensor product of channels, one has \cite{duan}
\[
S(\Phi_1\otimes\Phi_2)=S(\Phi_1)\otimes S(\Phi_2).
\]

A subspace $C\subseteq H_A$ is called an \emph{exact quantum code} for $\Phi$ if there exists a recovery quantum channel $\mathcal R$ such that
\[
\mathcal R\bigl(\Phi(\rho)\bigr)=\rho
\quad\text{for all }\rho\in\mathcal S(H_A),\quad \operatorname{supp}(\rho)\subseteq C.
\]
The dimension of the code is $\dim_{\C}C$; it equals the rank of the orthogonal projector $P$ onto $C$. By the Knill--Laflamme criterion \cite{knill}, exact recovery is equivalent to the condition

\[
PXP=\lambda_X P,\qquad \lambda_X\in\C\quad\forall X\in S(\Phi).
\]
For an arbitrary operator system $S$, a subspace $C=\Ran P$ satisfying $PSP=\C P$ is called a \emph{quantum anticlique} (or a code for $S$).

The maximum dimension among all such codes is denoted by
\[
\alpha_q(S)=\max\{\rank P:PSP=\C P\},
\]
where the maximum is taken over nonzero orthogonal projectors on $H$.
The one-shot zero-error quantum capacity is
\[
\overline{Q}_0(S)=\log_2\alpha_q(S).
\]
For a channel, set $\overline{Q}_0(\Phi)=\overline{Q}_0(S(\Phi))$. Superactivation of $\overline{Q}_0$ means that
\[
\overline{Q}_0(\Phi_1)=\overline{Q}_0(\Phi_2)=0,\qquad \overline{Q}_0(\Phi_1\otimes\Phi_2)>0.
\]

An example of this effect was given by M.~E.~Shirokov \cite{shirokov}. That work uses the eight-dimensional operator systems in $M_4$
\[
\mathcal N_\theta=\left\{
\begin{pmatrix}
a&b&e&f\\ c&d&f&\overline\gamma e\\
g&h&a&b\\ h&\gamma g&c&d
\end{pmatrix}:a,b,c,d,e,f,g,h\in\C\right\},
\qquad \gamma=e^{i\theta}.
\]
For channels $\Phi_\theta$ with these graphs, $\overline{Q}_0(\Phi_\theta)=0$ for $\theta\in(-\pi,\pi)$, whereas, when $\theta_1+\theta_2=\pi$, their tensor product has the two-dimensional code
\[
\spn\left\{\frac{\ket{11}+i\ket{22}}{\sqrt2},\
\frac{\ket{33}+i\ket{44}}{\sqrt2}\right\}.
\]
Here $\ket{1},\ldots,\ket{4}$ is the standard orthonormal basis of $\C^4$. Moreover, for positive $\theta_1,\theta_2,\theta_3$ whose sum is $\pi$, all pairwise tensor products have zero $\overline{Q}_0$, while the triple tensor product has positive $\overline{Q}_0$.

A symmetric example with input dimension $4$, in which $\overline{Q}_0(\Phi)=0$ but $\overline{Q}_0(\Phi\otimes\Phi)\ge1$, was also studied in detail by M.~E.~Shirokov and T.~V.~Shulman \cite{shirokovshulman}. The same work also showed that superactivation of two channels is impossible if one of them has a qubit input.

The aim of the present paper is to determine the minimum input dimension for which such superactivation is possible.

\section{Preliminary lemmas}
To prove additivity in the two-dimensional case, two simple properties of $\alpha_q(S)$ will be needed: monotonicity under inclusion of operator systems and its value for the tensor product of a scalar system and a diagonal system.

\begin{lemma}[Monotonicity under inclusion]\label{lem:mono}
Let $T\subseteq S$ be two operator systems on the same Hilbert space. Then
\[
\alpha_q(S)\le\alpha_q(T).
\]
\end{lemma}
\begin{proof}
If a projector $P$ satisfies the Knill--Laflamme condition for all $X\in S$, it also satisfies it for all $X\in T$. Thus every code for $S$ is a code for $T$.
\end{proof}

\begin{lemma}\label{lem:diag}
Let $m,r\ge1$ be integers, and let
\[
D_r=\{\diag(a_1,\ldots,a_r):a_j\in\C\}\subset M_r,
\]
\[
\C I_m=\{aI_m:a\in\C\}\subset M_m.
\]
Then
\[
\alpha_q(\C I_m\otimes D_r)=m.
\]
\end{lemma}
\begin{proof}
The lower bound is immediate. Fix a basis vector $\ket{j_0}\in\C^r$ and consider
\[
C=\C^m\otimes\spn\{\ket{j_0}\}.
\]
For any
\[
X=I_m\otimes\diag(a_1,\ldots,a_r)
\]
the compression of $X$ to $C$ equals $a_{j_0}I_C$. Therefore,
\[
\alpha_q(\C I_m\otimes D_r)\ge m.
\]
To prove the reverse bound, let $\{\ket{\psi_\ell}\}_{\ell=1}^M$ be an orthonormal basis of an arbitrary code for $\C I_m\otimes D_r$. Let $\{\ket j\}_{j=1}^r$ denote the standard orthonormal basis of $\C^r$. Write
\[
\ket{\psi_\ell}=\sum_{j=1}^r\ket{u_{\ell j}}\otimes\ket j,
\qquad \ket{u_{\ell j}}\in\C^m.
\]
For $\ell\ne s$, the Knill--Laflamme condition for $I_m\otimes\diag(a_1,\ldots,a_r)$ gives
\[
\sum_{j=1}^r a_j\langle u_{\ell j}|u_{sj}\rangle=0
\quad\forall a_1,\ldots,a_r,
\]
and hence
\[
\langle u_{\ell j}|u_{sj}\rangle=0\quad\forall j,\ \ell\ne s.
\]
Equality of the diagonal matrix elements gives
\[
\|u_{\ell j}\|^2=\|u_{sj}\|^2\quad\forall j,\ell,s.
\]
If $M>m$, then, for each fixed $j$, the $m$-dimensional space $\C^m$ contains $M$ pairwise orthogonal vectors $\{u_{\ell j}\}_{\ell=1}^M$; at least one of them must therefore be zero. Equality of their norms then implies that $u_{\ell j}=0$ for all $\ell$. This holds for every $j$, so all the vectors $\ket{\psi_\ell}$ must be zero, which is impossible. Thus $M\le m$.
\end{proof}

\section{Additivity for qubit inputs}
We first describe the possible values of $\alpha_q$ for systems in $M_2$ and identify a diagonal subsystem in each nonscalar system.
\begin{lemma}\label{lem:scalar}
Let $S\subset M_2$ be an operator system. Then
\[
\alpha_q(S)=2\quad\Longleftrightarrow\quad S=\C I_2.
\]
In all other cases, $\alpha_q(S)=1$.
\end{lemma}
\begin{proof}
If a two-dimensional code exists, its projector is $P=I_2$. The Knill--Laflamme condition then becomes
\[
X=\lambda_X I_2\quad\forall X\in S,
\]
that is, $S=\C I_2$. The converse is immediate. A one-dimensional code always exists.
\end{proof}

\begin{lemma}\label{lem:subsystem}
Let $S\subset M_2$ be an operator system other than $\C I_2$. Then $S$ contains a two-dimensional operator subsystem that coincides with $D_2$ after a unitary change of basis.
\end{lemma}
\begin{proof}
Choose $X\in S$ that is not proportional to the identity (that is, is nonscalar). Since $S=S^\dagger$, both Hermitian parts of $X$,
\[
H_1=\frac{X+X^\dagger}{2},\qquad H_2=\frac{X-X^\dagger}{2i}
\]
belong to $S$. At least one of them is nonscalar; denote it by $H$. A nonscalar Hermitian matrix $H\in M_2$ has two distinct eigenvalues $\lambda_1\ne\lambda_2$, so there exists a unitary operator $U$ such that
\[
U^\dagger HU=\begin{pmatrix}\lambda_1&0\\0&\lambda_2\end{pmatrix}.
\]
Since $\lambda_1\ne\lambda_2$, the linear span of $I_2$ and $U^\dagger HU$ is the entire diagonal algebra:
\[
U^\dagger\spn\{I_2,H\}U=D_2.
\]
\end{proof}

\begin{theorem}[Additivity for qubit channels]\label{thm:qubit}
Let $S_1,\ldots,S_n\subset M_2$ be the noncommutative graphs of channels with two-dimensional inputs. Then
\[
\overline{Q}_0(S_1\otimes\cdots\otimes S_n)=\sum_{i=1}^n \overline{Q}_0(S_i).
\]
\end{theorem}
\begin{proof}
Suppose that exactly $k$ of the systems $S_1,\ldots,S_n$ are scalar. A permutation of tensor factors is a unitary transformation, so, without loss of generality, assume that
\[
S_1=\cdots=S_k=\C I_2,\qquad S_{k+1},\ldots,S_n\ne\C I_2.
\]
When $k=0$ or $k=n$, use the conventions $(\C^2)^{\otimes0}=\C$ and $D_2^{\otimes0}=\C$.
We first obtain a lower bound. Fix an arbitrary unit vector
\[
\ket\eta\in(\C^2)^{\otimes(n-k)}
\]
and consider the subspace
\[
C=(\C^2)^{\otimes k}\otimes\spn\{\ket\eta\}.
\]
Its dimension is $2^k$. The operators in $S_1\otimes\cdots\otimes S_k=\C I_{2^k}$ act as scalars on the first $k$ tensor factors of $C$, and the compression of any operator to the one-dimensional space $\spn\{\ket\eta\}$ is also scalar. Thus the Knill--Laflamme conditions hold on $C$, and $C$ is a code for $S_1\otimes\cdots\otimes S_n$. Consequently,
\begin{equation}\label{eq:lower}
\alpha_q(S_1\otimes\cdots\otimes S_n)\ge2^k.
\end{equation}
For each $i>k$, the preceding lemma gives a unitary operator $U_i$ such that
\[
D_2\subseteq U_i^\dagger S_iU_i.
\]
Therefore, for some unitary $U$,
\[
\C I_{2^k}\otimes D_2^{\otimes(n-k)}\subseteq U^\dagger(S_1\otimes\cdots\otimes S_n)U.
\]
But
\[
D_2^{\otimes(n-k)}=D_{2^{n-k}}.
\]
Unitary conjugation of a noncommutative graph does not change $\alpha_q$, and Lemmas~\ref{lem:mono} and~\ref{lem:diag} give
\[
\alpha_q(S_1\otimes\cdots\otimes S_n)
\le\alpha_q(\C I_{2^k}\otimes D_{2^{n-k}})=2^k.
\]
Together with \eqref{eq:lower}, this yields
\[
\alpha_q(S_1\otimes\cdots\otimes S_n)=2^k.
\]
For each individual factor,
\[
\overline{Q}_0(S_i)=\begin{cases}1,&S_i=\C I_2,\\0,&S_i\ne\C I_2.\end{cases}
\]
Consequently,
\[
\overline{Q}_0(S_1\otimes\cdots\otimes S_n)=\log_2 2^k=k=\sum_{i=1}^n \overline{Q}_0(S_i).
\]
\end{proof}
\begin{corollary}
Superactivation of the one-shot zero-error quantum capacity is impossible for a pair of channels with two-dimensional input spaces.
\end{corollary}

\section{An explicit example with three-dimensional inputs}

We construct two noncommutative graphs $S,T\subset M_3$ such that
\[
\overline{Q}_0(S)=\overline{Q}_0(T)=0,
\qquad \overline{Q}_0(S\otimes T)\ge1.
\]
Fix the standard orthonormal basis $\ket0,\ket1,\ket2$ of $\C^3$ and write $E_{ij}=\ket i\bra j$. Set
\[
H_+=\begin{pmatrix}0&0&0\\0&1&1\\0&1&-1\end{pmatrix},
\qquad
H_-=\begin{pmatrix}0&0&0\\0&1&-1\\0&-1&-1\end{pmatrix}
\]
and define
\begin{equation}\label{eq:system}
\begin{aligned}
S&=\spn_{\C}\{I_3,E_{01},E_{10},H_+\},\\
T&=\spn_{\C}\{I_3,E_{02},E_{20},H_-\}.
\end{aligned}
\end{equation}
Both spaces contain the identity matrix and are closed under Hermitian conjugation, since $E_{ij}^\dagger=E_{ji}$ and $H_\pm^\dagger=H_\pm$. Thus $S$ and $T$ are operator systems. By the theorem on the realization of operator systems as noncommutative graphs of quantum channels \cite[Proposition~2]{reversibility}, there exist channels $\Phi_1,\Phi_2$ with input spaces $\C^3$ such that $S(\Phi_1)=S$ and $S(\Phi_2)=T$.

\subsection{Absence of two-dimensional anticliques for the individual graphs}
\begin{proposition}\label{prop:individual}
For the systems $S,T$ in \eqref{eq:system},
\[
\alpha_q(S)=\alpha_q(T)=1,
\qquad \overline{Q}_0(S)=\overline{Q}_0(T)=0.
\]
\end{proposition}
\begin{proof}
Suppose that $S$ has a two-dimensional code with orthogonal projector $P$. The Knill--Laflamme condition for $E_{01}\in S$ gives
\[
PE_{01}P=(P\ket0)(P\ket1)^\dagger=\lambda P.
\]
The left-hand side has rank at most one, whereas $\rank P=2$. Hence $\lambda=0$, and $P\ket0=0$ or $P\ket1=0$. Since $P$ has rank two in $\C^3$, its range coincides with $\ket0^\perp$ or $\ket1^\perp$, respectively.

In the first case, the compression of $H_+$ to $\Ran P$ in the basis $(\ket1,\ket2)$ is
\[
\begin{pmatrix}1&1\\1&-1\end{pmatrix},
\]
and in the second case, in the basis $(\ket0,\ket2)$, it is $\diag(0,-1)$. Both matrices are nonscalar, contradicting the Knill--Laflamme condition for $H_+\in S$.

Similarly, for a two-dimensional code of $T$, the condition for $E_{02}$ implies $\Ran P=\ket0^\perp$ or $\Ran P=\ket2^\perp$. The compressions of $H_-$ to these planes in the bases $(\ket1,\ket2)$ and $(\ket0,\ket1)$ are, respectively,
\[
\begin{pmatrix}1&-1\\-1&-1\end{pmatrix},
\qquad \diag(0,1).
\]
These are also nonscalar. Hence neither $S$ nor $T$ has a code of dimension greater than one.
\end{proof}

\subsection{Construction of a code for the tensor product}\label{subsec:product}
Choose the orthonormal vectors in $\C^3\otimes\C^3$
\[
\ket u=\ket{00},\qquad
\ket v=\frac{\ket{11}+\ket{22}}{\sqrt2},
\]
where $\ket{ij}=\ket i\otimes\ket j$. Set
\[
C=\spn\{\ket u,\ket v\},\qquad P=P_C.
\]
We show how the choice of generators in \eqref{eq:system} ensures the Knill--Laflamme conditions on $C$.

First, we verify that the off-diagonal matrix elements vanish. For arbitrary $A\in S$ and $B\in T$,
\[
\bra u A\otimes B\ket v
=\frac{A_{01}B_{01}+A_{02}B_{02}}{\sqrt2}.
\]
The generators $E_{01},E_{10}$ of the first graph and $E_{02},E_{20}$ of the second graph are chosen so that $A_{02}=0$ and $B_{01}=0$. Both terms therefore vanish. Similarly,
\[
\bra v A\otimes B\ket u
=\frac{A_{10}B_{10}+A_{20}B_{20}}{\sqrt2}=0,
\]
since $A_{20}=0$ and $B_{10}=0$.

To establish equality of the diagonal matrix elements, we use the lower blocks of $H_+$ and $H_-$,
\[
D_+=\begin{pmatrix}1&1\\1&-1\end{pmatrix},
\qquad
D_-=\begin{pmatrix}1&-1\\-1&-1\end{pmatrix}.
\]
They satisfy
\[
\operatorname{Tr}D_+=\operatorname{Tr}D_-=0,
\qquad \operatorname{Tr}(D_+^{\mathsf T}D_-)=0.
\]
The last equality is precisely what determines the choice of opposite signs for the off-diagonal entries of these blocks.

Let $A'$ and $B'$ denote the compressions of $A$ and $B$ to $\spn\{\ket1,\ket2\}$ in the indicated basis. By the definitions of the graphs,
\[
A'=aI_2+xD_+,
\qquad B'=bI_2+yD_-,
\qquad a=A_{00},\quad b=B_{00},
\]
where $a,b,x,y\in\C$. For the chosen vector $\ket v$, we obtain
\[
\begin{aligned}
\bra v A\otimes B\ket v
&=\frac12\operatorname{Tr}\bigl((A')^{\mathsf T}B'\bigr)\\
&=ab+\frac{ay}{2}\operatorname{Tr}D_-
     +\frac{bx}{2}\operatorname{Tr}D_+
     +\frac{xy}{2}\operatorname{Tr}(D_+^{\mathsf T}D_-)\\
&=ab=\bra u A\otimes B\ket u.
\end{aligned}
\]
Thus, for any $A\in S$ and $B\in T$,
\begin{equation}\label{eq:compression}
P(A\otimes B)P=A_{00}B_{00}P.
\end{equation}
By linearity,
\[
P(S\otimes T)P=\C P.
\]
Hence $C$ is a two-dimensional quantum anticlique, and
\[
\alpha_q(S\otimes T)\ge2,
\qquad \overline{Q}_0(S\otimes T)\ge1.
\]
Together with Proposition~\ref{prop:individual}, this gives
\[
\overline{Q}_0(S)=\overline{Q}_0(T)=0,
\qquad \overline{Q}_0(S\otimes T)\ge1.
\]
This constructs a pair of channels with input spaces $\C^3$ that exhibits superactivation of the one-shot zero-error quantum capacity.

\section{Minimum input dimension}
Define
\[
\begin{split}
d_{\min}=\min\bigl\{d:\ &\text{there exist channels }\Phi_1,\Phi_2\text{ with }\dim H_{A_i}\le d,\\
&\overline{Q}_0(\Phi_1)=\overline{Q}_0(\Phi_2)=0,\quad \overline{Q}_0(\Phi_1\otimes\Phi_2)>0\bigr\}.
\end{split}
\]
\begin{theorem}
The minimum input dimension for superactivation of the one-shot zero-error quantum capacity is
\[
d_{\min}=3.
\]
\end{theorem}
\begin{proof}
For $d\le2$, superactivation is impossible by Theorem~\ref{thm:qubit}: for any two channels with qubit inputs,
\[
\overline{Q}_0(\Phi_1\otimes\Phi_2)=\overline{Q}_0(\Phi_1)+\overline{Q}_0(\Phi_2).
\]

Consequently,
\[
d_{\min}\ge3.
\]

On the other hand, the operator systems $S,T\subset M_3$ constructed above are realized as the noncommutative graphs of channels $\Phi_1,\Phi_2$ with three-dimensional inputs, for which
\[
\overline{Q}_0(\Phi_1)=\overline{Q}_0(\Phi_2)=0,
\qquad \overline{Q}_0(\Phi_1\otimes\Phi_2)\ge1.
\]
This pair of channels satisfies the conditions in the definition of $d_{\min}$.

Therefore,
\[
d_{\min}\le3.
\]
Combining the two bounds gives $d_{\min}=3$.
\end{proof}

\section*{Note on the use of AI}
The pair of noncommutative graphs $S,T\subset M_3$ presented in this paper was obtained with the assistance of the ChatGPT GPT-6 Astra model. After the construction was obtained, all statements used in the proof, including the absence of two-dimensional anticliques for $S$ and $T$ and the validity of the Knill--Laflamme conditions for the constructed two-dimensional code in $S\otimes T$, were checked separately in exact algebraic form.

\renewcommand{\refname}{References}
\urlstyle{same}

\end{document}